\documentclass{svjour2}                    
\smartqed  
\usepackage{graphicx}
\usepackage{amsfonts}
\usepackage{amsmath}
\usepackage{amssymb}
\usepackage{graphicx}
\usepackage{color}
\usepackage{ulem}
\newcommand{\beq}{\begin{equation}}
\newcommand{\eeq}{\end{equation}}
\newcommand{\beqa}{\begin{eqnarray}}
\newcommand{\eeqa}{\end{eqnarray}}
\newcommand{\beal}{\begin{align}}
\newcommand{\enal}{\end{align}}
\newcommand{\Eq}[1]{Eq.(\ref{#1})}
\newcommand{\Ex}[1]{(\ref{#1})}

\newcommand{\la}{\langle}
\newcommand{\ra}{\rangle}

\newcommand{\nm}{\nonumber}
\begin{document}

\title{Molecular theory of \\
anomalous diffusion
}
\subtitle{Application to Fluorescence Correlation Spectroscopy}

\titlerunning{Anomalous diffusion}        

\author{Jean Pierre Boon         \and
            James F. Lutsko 
}


\institute{J.P. Boon \at
               Physics Department, CP 231,
               Universit\'e Libre de Bruxelles, 1050 -  Belgium\\
              \email{jpboon@ulb.ac.be}           
           \and
                J.F. Lutsko \at
              Physics Department, CP 231,
               Universit\'e Libre de Bruxelles, 1050 -  Belgium\\
              \email{jlutsko@ulb.ac.be}
              }

\date{Received: date / Accepted: date}

\maketitle

\begin{abstract}
The nonlinear theory of anomalous diffusion is based on particle
interactions giving an explicit microscopic description of diffusive processes
leading to sub-, normal, or super-diffusion as a result competitive effects 
between attractive and repulsive interactions. We present the explicit analytical
solution to the nonlinear diffusion equation which we then use to compute the
correlation function which is experimentally measured by correlation 
spectroscopy. The theoretical results are applicable in particular to the analysis
of fluorescence correlation spectroscopy of marked molecules in biological 
systems. More specifically we consider the case of fluorescently labeled lipids   
and we find that the nonlinear  correlation spectrum reproduces 
very well the experimental data indicating sub-diffusive 
molecular motion of lipid molecules in the cell membrane. 
 

\end{abstract}

\section{Classical diffusion and anomalous diffusion}
\label{intro}

   There are many systems observed in nature and in the laboratory where it seems  
 natural to use the language of diffusion, but where one finds that the space or time dispersion
 of the diffusing objects do not obey the classical diffusion equation, which is an indication that the 
 objects do not move "freely":  obstacles, time delays, interactions can modify their trajectories 
 in such a way that their mean squared displacement deviates from the classical linear law 
 $\la r^2 \ra \sim t$ and the Gaussian structure of the dispersion is deformed or replaced by 
  a different distribution. Such non-classical distributions with non-exponential behavior are 
  generally the signature of  {\textit {anomalous diffusion}} and the mean squared displacement 
  then follows a power law in the time variable: $\la r^2 \ra \sim t^{{\gamma}}$ with 
  $0 < {\gamma}\neq1$ (while ${\gamma}=1$ for regular diffusion).  
  Depending on whether ${\gamma} < 1$ or 
  ${\gamma} > 1$, one talks about {\textit {sub-diffusion}} when the particles are e.g.  
  delayed in their diffusive motion by the presence of obstacles or because of the structural
  complexity of the medium or because of molecular interactions, and {\textit {super-diffusion}} when 
  their motion is e.g. enhanced by concentration effects or by an external field. 
   
  Fundamental constraints in constructing a theory of anomalous diffusion are then needed to reproduce 
  a mean-squared displacement that exhibits power-law behavior as a function of time and the
fundamental demand for the existence of self-similar solutions, i.e. such that {\it all} moments scale similarly, $\la r^{2n} \ra \sim t^{n\gamma}$. This implies that the distribution should have the form $f(r,t) = t^{-\gamma/2} \phi(r/t^{\gamma/2})$ for some function $\phi(x)$ (as is the case for classical diffusion).  
    
  In Einstein's random walk model~\cite{einstein}, the jumps can extended to lengths greater than one with different probabilities including  rests (jumps of length zero) without affecting the diffusive nature of the process. Generally, diverse microscopic dynamics can give rise to "diffusion" phenomena at the macroscopic level, but the underlying mechanisms may be quite different; for instance the distinction should be made between {\textit {molecular diffusion}} of tagged particles which, while identical to the medium particles, are made observable by radioactive or fluorescent markers \cite{RNA_Diff,molecularD} and {\textit {tracer diffusion}}  where  experimentally one follows trajectories of distinguishable particles seeded  in an active medium ~\cite{sanchez,dogariu}.
    
  Various approaches for a general description of diffusive phenomena have been developed in the past few years. They can be divided into three classes: 
  
\noindent (i)  the {\textit{fractional Fokker-Planck equation}}  (FFPE) is based on the 
continuous time random walk  model with a power law ansatz for the distribution 
of the time delays in the motion of the particles \cite{FFPE_PhysToday}
and describes the phenomenology of sub-diffusion \cite{Metzler_Review,Barkai_Review};
\footnote{The analysis has been generalised to include super-diffusion by introducing a
similar power law ansatz for the distribution of particle displacements \cite{Abe}.}

\noindent (ii) the {\textit{fractional Brownian motion}}  uses a generalized
random walk model with correlations between particle displacements 
leading to a diffusion equation with classical structure, but with time-dependent coefficients 
 \cite{fBm1,fBm2};

\noindent (iii)  the {\textit{nonlinear Fokker-Planck equation}}  (NLFP) for anomalous diffusion,
which we use in the present analysis, is obtained from a generalization of Einstein's 
mean field equation for the random walk  by allowing the probability for a jump from one site to another to depend on the walkers concentration. When this microscopic dynamics is constrained by the demand for diffusive-like scaling solutions,  it is found~\cite{BoonLutsko,LutskoBoon8} that the jump probabilities must have the form of a power law $\sim f^{(\alpha - 1)}(r,t)$, where $f(r,t)$  is the probability that a random walker be at position $r$ at time~$t$. This approach provides a molecular theory of anomalous diffusion based on particle interactions  leading to sub- or super-diffusion (see Fig.1 in \cite{LutskoBoon13}) as a result of a balance between attractive and repulsive interactions.  
\footnote{In the restricted case that the walkers have equal probability to move in any direction, the NLFP reduces to the phenomenological {\textit{porous media equation}} \cite{Muskat}.}

In the next section, we review the nonlinear theory of anomalous diffusion and its
main result, the nonlinear Fokker-Plank equation (NLFP) whose solutions 
are given explicitly in Section \ref{solution} and further discussed in 
Sections~\ref{anom_diff}  and~\ref{NL2Dim}. Fluorescence correlation spectroscopy (FCS)
is an interesting light scattering method which can be used to measure molecular 
diffusion in biological systems and thereby detect anomalous diffusion. 
The FCS correlation spectrum is computed analytically (i) for the case of classical diffusion
in Section \ref{FCS} and (ii) for anomalous diffusion using the distribution 
function obtained from the NLFP solution in Section \ref{application} where the
analytical results are compared with experimental data for the case of lipid molecules 
diffusion in cell membranes.

\section{The generalized Fokker-Plank equation}
\label{NLFP}

Generalizing the jump probabilities $P_j$ in Einstein's master equation 
by introducing a functional dependence on the distribution functions $f (r, t)$
at the starting point $r-j$ and at the end point $r$ of the jump, we have 
\beq
\label{P(ff)}
P_j = {p}_j\,{F}[f(r - j\,\delta r, t), f (r, t)]\,,\;\;\;\;\;\mbox{with}\;\;\;\;\sum_{j}\,\,P_j = 1\,,
\eeq
where the probabilities ${p}_j$ are drawn from a prescribed distribution and the bounding 
condition $0\leq P_j \leq1$ imposes $0\leq  F(x,y) \leq1$ as well as restrictions on the 
functional form of $F(x,y) \doteqdot {F}[f(r - j\,\delta r, t), f (r, t)]$. 
Under these conditions, multiscale expansion of the master 
equation is shown to give the generalized Fokker-Planck (or generalized diffusion) 
equation \cite{LutskoBoon13}
\begin{align}
&\frac{\partial f}{\partial t}+M_1\frac{\partial}{\partial r}\left[  xF\left(
x,x\right)  \right]  _{f}  \\
&= M_2\frac{\partial}{\partial r}\left[
\frac{\partial xF\left(  x,y\right)  }{\partial x}-\frac{\partial xF\left(
x,y\right)  }{\partial y}\right]  _{f} \frac{\partial f}{\partial
r}\nonumber\label{main2} \\ \nm
&+\frac{1}{2}M_1^{2}\delta t\frac{\partial}{\partial r}\left[  \frac{\partial
xF\left(  x,y\right)  }{\partial x}-\frac{\partial xF\left(  x,y\right)
}{\partial y}-\left(  \frac{\partial xF\left(  x,x\right)  }{\partial x}%
\right)  ^{2}\right]  _{f}\frac{\partial f}{\partial r}\,,
\end{align}
\label{GFPE}
with the notation 
\begin{equation}
\left[  \frac{\partial xF\left(  x,y\right)  }{\partial x}\right]
_{f}=\left[  \frac{\partial xF\left(  x,y\right)  }{\partial x}\right]
_{x=f\left(  r,t\right)  ,y=f\left(  r,t\right)  }\,.
\end{equation}
{{In \Eq{GFPE}, $M_1$ and $M_2$ are given by }}
\begin{align}
&M_1  = \frac{\delta r}{\delta t}\,\sum_{j}\,j\,p_{j} = \frac{\delta r}{\delta t}\,J_{1}\,,\nonumber\\
&M_2=\frac{1}{2}\frac{\left(  \delta r\right)^{2}}{\delta t}\left( \left(\sum_{j}\,j^2\,p_{j}\right) - %
J_{1}^{2}\right)  = \frac{1}{2}\frac{\left(  \delta r\right)^{2}}{\delta t}\left(  J_{2}-J_{1}^{2}\right) 
\end{align}
{{where the $J_{n}$'s denote the moments $J_{n}=\sum_{j}j^{n}p_{j}$. }}
Note that for $F\left( x,y\right) = 1$, \Eq{GFPE} reduces to the classical 
advection-diffusion equation.

Since the function $F\left(  x,y\right)$ is defined in terms of the jump probabilities, 
it must be bounded, and  so must satisfy 
\begin{equation}
0\leq xF\left(  x,y\right)  \leq1\;\;\;\;\mbox{and}\;\;\;\;0\leq yF\left(
x,y\right)  \leq1\,,\;\;\;\;\forall \;x,y\in\left[  0,1\right]\,. 
\label{bounds}
\end{equation}
It was shown in \cite{LutskoBoon13} that self-similar solutions are possible if and only if 
\begin{equation}
\lim_{y \rightarrow x} \left[\frac{\partial xF\left(  x,y\right)  }{\partial x}-\frac{\partial xF\left(
x,y\right)  }{\partial y}\right] \sim x^{\alpha-1}
\end{equation}
where the scaling exponent $\alpha$ is related to the diffusion exponent by
\begin{equation}
\gamma=\frac{2}{\alpha + 1}\;.
\end{equation}

Anomalous (sub- and super-) diffusion can be described in a single 
formulation when the jump probabilities have the following form in terms
of the occupation probabilities 
\begin{equation}
\label{ansatz}
{F}(x,y; \omega_s, \omega_e) \sim \omega_s\,F_s(x) + \omega_e \,F_e(y) \,;
\end{equation}
here $\omega_s$ and $\omega_e$ are weighting factors relative to the functionals 
of the concentrations at the starting point and at the end point of the jump. 
Using the notation $a \equiv \omega_e/\omega_s$ ,  \Ex{ansatz} is rewritten in 
normalised form as
\begin{equation}
F(x,y;a) = \frac{F(x) + aF(y)}{F(x)+F(y)}.
\label{model_alt}
\end{equation}
where the positivity arguments, $x,y$ and the constraints (\ref{bounds}) and $F(x,y;a) \ge~0$ 
imply that $0 \le a \le 1$. 

Considering the case that there is no drift ($M_1=0$ in \Eq{GFPE}), and in order that  
the general formulation describe diffusion, we should have a scaling solution of the form
$f\left(  r,t\right)  =t^{-\gamma/2}\phi\left(  r/t^{\gamma/2}\right)$ which demands that
 \cite{LutskoBoon13}
\begin{equation}
\lim_{y\rightarrow x}\left( \frac{\partial }{\partial x}xF\left( x,y; a \right) -%
\frac{\partial }{\partial y}xF\left( x,y; a, \alpha\right) \right) = K\, x^{\alpha - 1 } \,,
\label{scaling_cond}
\end{equation}%
for some constant $\alpha$. Using  \Ex{model_alt} in the l.h.s of \Ex{scaling_cond} gives
\begin{align}
\frac{1 + a}{2} + \frac{1-a}{2} \frac{x\,F^{\prime }\left( x\;a,\alpha \right) }{\,F\left(x;a,\alpha\right) }& = K\,  x^{\alpha - 1 }  \,,
\end{align}%
which is solved to yield
\begin{align}
F\left( x;a,\alpha\right)& =\frac{B}{x^\frac{1+a}{1 -a}}\,\exp {\left( \frac{2\,K  }%
{1-a}\,\frac{x^{\alpha - 1 }}{\alpha - 1}\right) } \,,
\label{fx} 
\end{align}%
where $B$ is an integration constant; reinserting \Ex{fx}  into  \Ex{model_alt}, we find
\begin{align}
F(x,y; a, \alpha) = \frac{1 + a \,\left(\frac{x}{y}\right)^{\frac{1+a}{1-a}}\,\exp \left( \frac{2\,K} %
{1-a}\,\frac{y^{{\alpha -1}}-x^{{\alpha -1}}}{\alpha -1}\right) }%
{1+  \left(\frac{x}{y}\right)^{\frac{1+a}{1-a}}\,\exp \left( \frac{2\,\lambda  } %
{1-a}\,\frac{y^{{\alpha -1}}-x^{{\alpha -1}}}{\alpha -1}\right)} \,.
\label{sol_a}
\end{align}%

The natural limit: $\lim_{{{\alpha \rightarrow 1}}}F(x,y;a)=1$ requires { $ K= \frac{1}{2}(1 + a)\, \lambda  ^{\alpha-1}$ where $\lambda  $ is an unitary constant with the dimension of length.  Thus,}
\begin{align}
F(x,y; a, \alpha) = \frac{1 + a\, G(x,y; a, \alpha)}{1 +  G(x,y; a, \alpha)} \,,
\label{F_G}
\end{align}%
with{
\begin{align} 
{G(x,y; a, \alpha)} \,=\,
\left(\frac{x}{y}\right)^{\frac{1+a}{1-a}}\,\exp \left( \frac{1+a}{1-a}\; %
\lambda^{\alpha-1} \,\frac{y^{{\alpha -1}}-x^{{\alpha -1} }}{\alpha -1} \right)\;.
\label{sol_F_alt} 
\end{align}%
It is clear that for any finite value $x>0$, $0\leq F(x,y;a, \alpha)<1 \;\;, \forall \; y \in \left[0,1\right]\,$  
and that the limit $F(x,y;a=1, \alpha\rightarrow 1)=~1$ gives normal diffusion. Furthermore%
\begin{equation} \label{zerox}
\lim_{x\rightarrow 0}F(x,y; a, \alpha)=\left\{ 
\begin{array}{c}
1,\;\;\;\alpha > 1 \\ 
a,\;\;\;\alpha < 1%
\end{array}%
\right.
\;\;\;\mbox{;}\;\;
\lim_{y\rightarrow 0}F(x,y; a, \alpha)=\left\{ 
\begin{array}{c}
a,\;\;\;\alpha > 1 \\ 
1,\;\;\;\alpha < 1  %
\end{array} 
\right.
\end{equation}

\paragraph{Physical interpretation.} To provide some interpretation, we note that%
{
\begin{align}
\frac{\partial }{\partial x}F(x,y; a, \alpha)  
& = (1+a)\, \frac{G(x,y; a, \alpha)}{\left({1 +  G(x,y; a, \alpha)}\right)^2}\,\left( \frac {(\lambda \, x)^{\alpha - 1} - 1}{x}\right)\;,
\end{align}
\begin{align}
\frac{\partial }{\partial y}F(x,y; a, \alpha)   
& = (1+a)\, \frac{G(x,y; a, \alpha)}{\left({1 +  G(x,y; a, \alpha)}\right)^2}\,\left( \frac {1-(\lambda \, y)^{\alpha - 1}}{y}\right)\;.
\end{align}
Since $0<x,y<1$, the signs of these derivatives are determined by the factors on the right:{
$\left( (\lambda\,  x)^{\alpha - 1} - 1\right)$ and $\left( 1 - (\lambda \, y)^{\alpha-1} \right) $}, and so depend on whether 
($\alpha -1$)  is positive or negative:
\begin{align}
\alpha & > 1\Longrightarrow \frac{\partial }{\partial x}F(x,y; a, \alpha)<0<\frac{\partial 
}{\partial y}F(x,y; a)  \notag \\
\alpha & < 1\Longrightarrow \frac{\partial }{\partial x}F(x,y; a, \alpha)>0>\frac{\partial 
}{\partial y}F(x,y; a)  
\end{align}%

In the first case, $\alpha  > 1$, the jump probability decreases with the concentration
at the starting point and increases with the concentration at the arrival site; in other
words the jump rate is reduced by putting more walkers at the origin and increased 
by putting more at the terminus of a jump: this is analogous to an attractive interaction. 
For $\alpha < 1$, we have the reverse situation: the jump rate is increased by putting 
more walkers at the origin and decreased by putting more at the terminus, thus 
emulating a repulsive interaction. In the standard problem with all walkers at the origin 
at $t=0$, the distribution decays monotonically away from the origin; thus, if the particles 
repel, the distribution expands faster (i.e. tends to a uniform distribution more
quickly) whereas if they attract, then this attraction slows down the spread of the distribution.
The physical interpretation is that  attractive interactions give sub-diffusion and 
repulsive interactions give super-diffusion.

\section{Solutions of the nonlinear diffusion equation}
\label{solution}

{{In the absence of drift, the generalized Fokker-Plank equation, Eq.(\ref{GFPE}) with \Ex{F_G} 
and \Ex{sol_F_alt}, gives the nonlinear diffusion equation \cite{Muskat}
\begin{equation}
\frac{\partial f(r, t)}{\partial t}=%
\lambda  ^{\alpha-1} \frac{1+a}{2\,\alpha}\,M_2 \frac{\partial^2 }{%
\partial r^2}\,f^{\alpha}(r, t) \,,
\label{GADE1}
\end{equation}%
(here $M_2 = \frac{1}{2}\frac{\left(  \delta r\right)^{2}}{\delta t}\,\sum_{j}\,j^2\,p_{j}$)
and the scaling solutions are obtained following the development given in \cite{LutskoBoon13}
yielding
\begin{equation}
f\left( r, t\right) =t^{-\gamma /2}\,W\left( 1\pm V\,\frac{r^{2}}{t^{\gamma }}\right) ^{1/(\alpha-1)}\,,
\label{sol_GADE1}
\end{equation}%
 where $W$ and $V$ are determined by the normalization condition (see below) and by the expression 
 obtained  by inserting \Ex{sol_GADE1} into \Ex{GADE1}
\begin{equation}
V W^{\alpha -1} = \lambda  ^{1-\alpha}\,\frac{|1-\alpha|}{1+\alpha}\,\frac{1}{1+a}\,{M_2}^{-1}\,.
\label{VW1}
\end{equation}%

 \paragraph{(i) Super-diffusive case.} For $\alpha <1$, %
\begin{equation}
f\left( r, t\right) =t^{-\gamma /2}\,W\left( 1+V\,\frac{r^{2}}{t^{\gamma }}\right) ^{1/(\alpha-1)}\,.
\end{equation}
The normalization condition (using the reduced variable $\zeta = V^{1/2}\,\frac{r}{t^{\gamma/2}}$) reads
\begin{equation*}
W\,V^{ -1/2}\,\int_{-\infty}^{\infty }\,d\zeta\,\left( 1+\zeta^{2}\right) ^{1/(\alpha-1) } =1  \;\;\;\Longrightarrow\;\;\;
\frac{W}{\sqrt V} = \frac {1}{\sqrt \pi} \, \frac{\Gamma(\frac{1}{1-\alpha})}{\Gamma(\frac{1}{1-\alpha} - \frac{1}{2})} \,,
\end{equation*}%
 provided 
\begin{equation}
\frac{\alpha +1}{\alpha-1 }<0\;\;\;\Longrightarrow\;\;\; -1<\alpha <1\, \;\;\;\Longrightarrow\;\;\; 
\gamma > 1 \,, %
\end{equation}%
and the mean-squared displacement 
$\left\langle r^{2}\right\rangle  =\int_{-\infty }^{\infty }r^{2}f\left(r;t\right) dr$ is given by
\beqa
\left\langle r^{2}\right\rangle &=&t^{\gamma }\,W\,V^{-3/2} %
\int_{-\infty }^{\infty } d\zeta \, \zeta^{2}\left(1+\zeta^{2}\right) ^{1/(\alpha-1) } \nm  \\
&=& \,\frac{W}{\sqrt V} \,\frac{t^{\gamma }}{V}\,\frac{\sqrt \pi}{2}%
\frac{\Gamma(\frac{3\alpha - 2}{2(1-\alpha)})}{\Gamma(\frac{1}{1-\alpha})} %
\,=\,\frac{t^{\gamma }}{V} %
 \, \frac{\Gamma(\frac{3\alpha - 2}{2(1-\alpha)})}{2\,\Gamma(\frac{1+\alpha}{2(1-\alpha)})} \,,
\label{msd_super}
\eeqa%
which is finite if 
\begin{equation} \label{bound}
\frac{3\alpha - 1}{\alpha-1 }<0\;\;\;\Longrightarrow\;\;\; \frac{1}{3}<\alpha <1\;\;\;\Longrightarrow\;\;\; 
\frac{3}{2} > \gamma >1 \,.
\end{equation}

\paragraph{(ii) Sub-diffusive case.} For $0< \gamma < 1$ i.e. $\alpha >1$ 
the distribution has finite support so that 
\begin{equation} 
f\left( r, t\right) =t^{-\gamma /2} \,W\left( 1- V\,\frac{r^{2}}{t^{\gamma }%
}\right) ^{1/(\alpha-1) }\Theta \left( 1- V\,\frac{r^{2}}{t^{\gamma }}\right) \,,
\label{dist1}
\end{equation}%
with the normalization condition%
\begin{equation*} 
W\,V^{ -1/2}\,\int_{-1}^{1}\,d\zeta\,\left( 1 - \zeta^{2}\right) ^{1/(\alpha-1) } =1 \;\;\;\Longrightarrow\;\;\;
\frac{W}{\sqrt V} = \frac {1}{\sqrt \pi} \, \frac{\Gamma (\frac{2\alpha}{\alpha -1}  + 1)}{\Gamma (\frac{1}{\alpha -1} + 1)} \,,
\end{equation*}%
and the mean-squared displacement reads%
\beqa
\left\langle r^{2}\right\rangle &=&t^{\gamma }\,W\,V^{-3/2} %
\int_{-1}^{1} d\zeta \zeta^{2}\left(1- \zeta^{2}\right) ^{1/(\alpha-1) } \nm \\ 
&=& \,\frac{W}{\sqrt V} \,\frac{t^{\gamma }}{V}\, \frac{\sqrt \pi}{2}\,
\frac{\Gamma(\frac{1}{\alpha -1} +1)}{\Gamma(\frac{1}{\alpha-1} +\frac{5}{2})} %
\,=\,\frac{t^{\gamma }}{V} \,
 \, \frac{\Gamma(\frac{3\alpha -1}{\alpha -1})}{2\,\Gamma(\frac{5\,\alpha - 3}{\alpha -1} )} \,.
\label{msd_sub}
\eeqa%

\section{The anomalous diffusion coefficient}
\label{anom_diff} 

The continuum results  imply that an initial distribution of the form 
$f_{0}(r) = W^{\prime}\,(1+s_{\alpha }\left( r/w\right) ^{2})^{1/(\alpha-1) }_{+}$
with $W^{\prime}$ determined by normalization and $s_{\alpha} = \mp$ for $\alpha \gtrless 1$, 
will evolve self-similarly with mean-squared displacement increasing as $t^{\gamma }$. 
Indeed determining the quantities $W$ and $V$ by combining the normalization condition with
\Ex{VW1}, the mean-squared displacement for both sub- and super-diffusion, takes the form 
\begin{equation}
\left\langle r^{2}\right\rangle = \widetilde{D}_{\gamma} \, t^{\gamma } \,,
\label{msd}
\end{equation}
with
\begin{equation}
 \widetilde{D}_{\gamma} = {\mbox{const}}\times  W\,V^{-3/2} = 
{\mbox{const}} \times \lambda  ^{2(1- \gamma )}\,\left(%
\frac{1+a}{ \gamma - 1} \right)^{\gamma}\,{M_2}^{\gamma}\,.
\label{D_gamma}
\end{equation}
$\widetilde{D}_{\gamma}$ is the {\it anomalous  diffusion coefficient} and has dimensions 
$[\widetilde{D}_{\gamma}] = L^2 \, T^{-\gamma}$.  
The distribution function then reads explicitly for super-diffusion (${\alpha} < 1$)
\begin{equation}
f\left( r, t\right) = \,\frac{1}{\sqrt {m_{\alpha}\,\pi {\widetilde{D}_{\gamma}\,{t^{\gamma}}}}}
\left( 1 + \,\frac{r^{2}}{m_{\alpha}\,\widetilde{D}_{\gamma}\,t^{\gamma }}\right) ^{1/(\alpha-1)} \,,
\label{sol_GADEsuper}
\end{equation}%
and for sub-diffusion (${\alpha} > 1$)
\begin{equation}
f\left( r, t\right) = \,\frac{1}{\sqrt {n_{\alpha}\,\pi {\widetilde{D}_{\gamma}\,{t^{\gamma}}}}}
\left( 1- \,\frac{r^{2}}{n_{\alpha}\,\widetilde{D}_{\gamma}\,t^{\gamma }}\right) ^{1/(\alpha-1)}\,
\Theta \left( 1-  \,\frac{r^{2}}{n_{\alpha}\,\widetilde{D}_{\gamma}\,t^{\gamma }}\right) \,,
\label{sol_GADEsb}
\end{equation}%
where  $ m_{\alpha}$ and  $n_{\alpha}$ are constants. 

Similarly \Eq{GADE1} can also be written as
\begin{equation}
\frac{\partial }{\partial t} f(r, t)= \, \frac{\partial }{\partial r} {\cal{J}}_{\alpha}(r, t)\,%
 \;\;\;\;\;\mbox{with} \;\;\;\;\; {\cal{J}}_{\alpha}(r, t)\,=\,%
{\cal{D}}_{\alpha}\,\frac{\partial }{\partial r} f(r, t) \,,
\label{GADE3}
\end{equation}%
where ${\cal{J}}_{\alpha}(r, t)$ is the current density and here
${\cal{D}}_{\alpha} = \frac{1+a}{2}\,\left(\lambda  \,f \right)^{\alpha-1} M_2$ 
has the usual dimensions of a diffusion coefficient ($ L^{2} \, T^{-1}$). 
It follows that we have the relation 
$ \widetilde{D}_{\gamma}\,=\,c_{\gamma}\,{\cal{D}}_{\alpha}^{\gamma}$,
where $c_{\gamma}$ is a constant with $\lim_{\gamma \rightarrow 1} \, c_{\gamma}\, =1$.

These results emphasize that the anomalous diffusion coefficient $\widetilde{D}_{\gamma}$
cannot be defined in the usual sense $\lim_{t\rightarrow \infty} \frac{\la r^{2}(t)\ra}{t} = D$ 
(which would give the unphysical values $D=0$ for sub-diffusion and $D=\infty$ for super-diffusion)
but can be defined as a diffusion coefficient with fractional time dimension. In practice 
$\widetilde{D}_{\gamma}$ is evaluated from the mean squared displacement \Ex{msd} 
as measured experimentally \cite{RNA_Diff,molecularD,sanchez,dogariu} or as obtained 
by numerical simulation of the master equation \cite{LutskoBoon13}, both methods
giving a physically observable quantity.
Only in the limit ${\gamma = \alpha =1}$  and $a=1$ does one have the  the classical result:
$\widetilde{D}_{\gamma =1}\,=\,{\cal{D}}_{\alpha =1}\,=\,M_2 $  
 with dimension $ L^2 \, T^{-1}$. {Conversely we should note that the advection term in the 
 generalized Fokker-Planck equation reads ${\cal{C}}_{\alpha}\, \frac{\partial }{\partial r}f$ 
 where  ${\cal{C}}_{\alpha}\,=\, \frac{1+a}{2}\,\left({\lambda  }\,f\right)^{\alpha -1}\,M_1$ (with 
$M_1= \frac{\delta r}{\delta t}\,\sum_{j}\,j\,p_{j}$) has dimensions $[{\cal{C}}_{\alpha}] = L \, T^{-1}$. 
In the limit   $\alpha \rightarrow 1$ and $a=1$, one has  the usual advection-diffusion equation
and $f(r, t)$ takes the classical Gaussian form $f\left( r, t\right) \sim%
\exp \left(- \,\frac{(r - ct)^{2}}{4\,{D}\,t}\right)$ with $D = {\cal{D}}_{\alpha =1}= M_2$
and  ${c=\cal{C}}_{\alpha =1}\,=\,M_1$}.

\section{Nonlinear diffusion in $d$-dimensions}
\label{NL2Dim}

Considering molecular diffusion in a  $d$-dimensional
volume, the  nonlinear diffusion equation \Ex{GADE1} (for the sub-diffusive case $\alpha>1$)
\begin{equation}
\frac{\partial f(r, t)}{\partial t}= D_{\alpha}\,\nabla^{2}f^{\alpha}(r, t)\,,
\label{GADE2}
\end{equation}
with $D_{\alpha}\,=\lambda  ^{\alpha-1} \frac{1+a}{2\,\alpha}\,M_2$ ,
becomes, in $d$-dimensions with spherical symmetry,
\begin{equation}
\frac{\partial}{\partial t}f(r, t)=D_{\alpha}\,\frac{1}{r^{d-1}}\frac{\partial}{\partial r}%
r^{d-1}\frac{\partial}{\partial r}\,f^{\alpha}(r, t)=D_{\alpha}\,\left(  \frac{\partial^{2}}
{\partial r^{2}}\,+\,\frac{d-1}{r}\,\frac{\partial}{\partial r} \right)\,f^{\alpha}(r, t) \,.
\label{NLE2dim}
\end{equation}
A scaling solution will have the form%
\begin{equation}
f(r, t)=\frac{1}{t^{d\gamma}}\phi\left(  r/t^{\gamma/2}\right)  \equiv\frac
{1}{t^{d\gamma}}\phi\left(  \xi\right) \,,
\label{scaling}
\end{equation}
which gives
\begin{align}
\frac{\partial}{\partial t}f(r, t) %
 &  =-\frac{\gamma}{2}\,\frac{1}{{t^{1+d\gamma/2}}}\,(d\,\phi +\xi\phi^{\prime}) \,, \nm \\
\frac{\partial}{\partial r}f^{\alpha}(r, t)  %
&  =\frac{\alpha}{t^{\left( 1+d\,\alpha \right)  \gamma/2}}\,\phi^{\prime}\phi^{\alpha-1}\,,    \nm \\
\frac{\partial^{2}}{\partial r^{2}}f^{\alpha}(r, t)  %
&  =\frac{\alpha}{t^{\left(1+ d\,\alpha/2 \right)  \gamma}}\,\phi^{\alpha-2}%
\left(\phi\,  \phi^{\prime\prime}\,+ \,\left(\alpha-1\right)  \phi^{\prime2} \right) \,.
\label{terms}
\end{align}
Inserting these results into \Eq{NLE2dim}, it follows that in $d$-spherical dimensions 
there is a general relation between the anomalous exponent and the 
nonlinear exponent  
\begin{equation}
\frac{d\,\gamma}{2} + 1 = \frac{d\,\alpha\,\gamma}{2} + \gamma  \;\;\; \Longrightarrow \;\;\; 
\gamma= \frac{2}{2 + (\alpha -1) \,d} \;,%
\end{equation}
 that is
 \begin{equation}
 \mbox{in $1-d$} : \gamma = \frac{2}{1+ \alpha} \;\;;\;\;
\mbox{in $2-d$} : \gamma= 1/ {\alpha}\;\;;\;\;
\mbox{in $3-d$} : \gamma = \frac{2}{3\alpha -1}
\end{equation}
 
In the next sections we will consider an experimental situation in planar symmetrical
dimension in which case,  the scaling equation \Ex{NLE2dim}-\Ex{terms} 
 becomes
\begin{equation}
\alpha^{2}D\frac{d}{d\xi}\left(  \xi\phi^{\prime}\phi^{\alpha-1}%
+\frac{1}{2\alpha^{2}D}\xi^{2}\phi\right)\,=\,0 \,,
\end{equation}
which has the solution
\beq
\phi   =  B\left(  1-\frac{\alpha-1}{4B^{\alpha-1}D_{\alpha}\,\alpha^{2}}\xi^{2}\right)^%
{\frac{1}{\alpha-1}}\,,\;\;\;\mbox{with} \;\;\; \xi^{2}=\frac{r^{2}}{t^{1/\alpha}} \;.
\eeq
Evaluating the normalisation constant $B$, we finally obtain
\begin{equation}
f(r, t)\,=\,\left(  \frac{1}{4\pi\alpha D_{\alpha}\,t}\right)  ^{1/\alpha}\left(  1-\pi\frac
{\alpha-1}{\alpha}\frac{r^{2}}{\left(  4\pi\alpha D_{\alpha}\,t\right)  ^{1/\alpha}%
}\right)  _{+}^{\frac{1}{\alpha-1}} \;.
\label{2dSol}
\end{equation}

\section{The fluorescence correlation spectrum}
\label{FCS}

Fluorescence Correlation Spectroscopy (FCS) \cite{SchwilleHaustein,Elson} is an experimental 
technique by which one observes and records temporal changes in the fluorescence emission intensity caused by single fluorophores passing through the detection volume. The measured spectrum contains  the correlation function of the temporal fluctuations of fluorescently marked particles thereby providing a quantitative evaluation of their diffusing properties. The method is particularly appropriate for the study 
of biological molecules in their proper environment such as cells and cell membranes because 
the measurements  can be performed in very small volumes with a $\mu$m detection accuracy and 
at very low intensity illumination. %
From the analytical viewpoint, the application of the theory of anomalous diffusion to FCS is very
interesting because the computation of the fluorescence correlation spectrum involves the distribution function of the diffusing objects. Therefore in contrast to the simple typical measurement of the mean squared  displacement, FCS offers a possible measurable indication of different molecular mechanisms of diffusion. 

The fluorescence correlation signal $J(\tau)$ results from the convolution of the instrumental
form factor  $\cal F$ with the correlation function $\Phi$ of the diffusing particles with mean 
concentration $\la C \ra$ in the illuminated volume $V$ :
\begin{equation}
J(\tau) = I_{0} \, \left\langle C\right\rangle \int_V \Phi \left(\mathbf{r}%
_{1},\mathbf{r}_{2},\tau\right)  {\cal{F}}(r_1, r_2) \,d\mathbf{r}_{1}d\mathbf{r}_{2} \,,
\end{equation}
where $ I_{0}$ is the illumination intensity. The form factor is well approximated by a 
Gaussian  distribution over the detection volume of width $w$: 
${\cal{F}}(r_1, r_2) = exp[-(r_1^2 + r_2^2)/w^2]$, and 
$\Phi \left(\mathbf{r}_{1},\mathbf{r}_{2},\tau\right)$ describes the decay of concentration 
fluctuation correlations. Assuming $\Phi$ depends only on the distance between the fluctuations, 
i.e. $\Phi \left(  \mathbf{r}_{1},\mathbf{r}_{2},\tau\right)  =\Phi \left(  \left\vert \mathbf{r}%
_{1}-\mathbf{r}_{2}\right\vert ^{2},\tau\right)  \equiv \Phi \left(  r^{2},\tau\right) $ and defining
${R}  {\equiv} \vert \mathbf{R} \vert = \frac{1}{2}  \vert \mathbf{r}_{1}+\mathbf{r}_{2}\vert$ 
so that $r_{1}^{2}+r_{2}^{2}=2R^{2}+\frac{1}{2}r^{2}$, we obtain
\begin{equation}
J(\tau) = I_{0} \,\left\langle C\right\rangle \int_V \Phi \left(  r^{2},\tau\right)
e^{-2R^{2}/w^{2}}e^{-r^{2}/2w^{2}}d\mathbf{R}d\mathbf{r} \,,%
\end{equation}
which, when  the detection volume is  $2$-dimensional 
(such as e.g. in cell membranes), gives %
\begin{align}
J(\tau) & = I_{0} \, \left\langle C\right\rangle \left(  \int_{-\infty}^{\infty
}e^{-2x^{2}/w^{2}}dx\right)  ^{2}2\pi\int_{0}^{\infty} \Phi \left(  r^{2}%
,\tau\right)  e^{-r^{2}/2w^{2}}rdr  \nm \\
&  = I_{0} \left\langle C\right\rangle  \,\pi^{2}w^{2} \int_{0}^{\infty}
\Phi \left(  r^2, \tau\right)  e^{-r^2/2w^{2}} rdr \,.
\label{spectrum}
\end{align}

For classical diffusion, $\Phi$ is a Gaussian distribution and the fluorescence 
correlation spectrum is given by
\begin{align}
 J(\tau) & = I_{0} \, \left\langle C\right\rangle  \pi^{2}w^{2} %
 \int_{0}^{\infty} \left( e^{-  r^{2}/4D\tau} \right) e^{-r^{2}/2w^{2}}rdr \nm \\
&  =  I_{0} \left\langle C\right\rangle \frac{1}{2}\pi w^{2}\left(1+\frac{2D\tau}{w^{2}}\right)^{-1} \,.
\label{GausSpec}
\end{align}

\section{Anomalous molecular diffusion  in membranes}
\label{application}

In many instances, diffusion processes in biological systems do not obey
the classical description because particle diffusive motion is usually
hindered in crowed biological media often leading to sub-diffusion. 
When this is the case, the distribution, instead of the classical Gaussian, has a power
law structure as described in section \ref{solution}, and in the FCS analysis
of $2-d$ molecular diffusion in cell membranes \cite{Webb} the function $\Phi$ in 
\Ex{spectrum} must be the two-dimensional distribution  \Ex{2dSol} which gives 
\begin{equation}
J(\tau)   =  I_{0}\left\langle C\right\rangle \frac{\pi}{2} w^{2}\, {\cal{J}}(\tau) \,,
\end{equation}
with
\begin{align}
 &{\cal{J}}(\tau)   =  2\pi \,\int_{0}^{\infty} f(r, \tau)\;e^{-r^{2}/2w^{2}}rdr \nm \\
 &  = 2\pi  \int_{0}^{\infty
}\left(  \frac{1}{4\pi\alpha D_{\alpha}\tau}\right)  ^{1/\alpha}\left(  1-
\pi \frac{\alpha-1}{\alpha}\frac{r^{2}}{\left(  4\pi\alpha D_{\alpha}\tau\right)
^{1/\alpha}}\right)  _{+}^{\frac{1}{\alpha-1}}e^{-r^{2}/2w^{2}}rdr \,. \nm 
\end{align}
With  a change of variables 
\beq
\chi  = 1- \frac{\pi (1- 1/ \alpha) r^2}{\left(4\pi\, \alpha\,D_{\alpha}\,\tau\, \right)^{1/\alpha}} \,,
\eeq
and defining
\beq
K =\frac{\left(4\pi \,\alpha\,D_{\alpha}\tau \right)^{1/ \alpha}}{2 \pi (1-1/ \alpha)\,w^{2}} \,,
\label{defK}
\eeq
we have
\beq
{\cal{J}}(\tau) = \, \frac{\alpha}{\alpha -1}\, e^{- K}\,%
 \int_{0}^{1} \chi ^{\frac{1}{\alpha - 1}}\,e^{K\,\chi}\,d\chi \,.  
 \label{J_chi}
 \eeq
 
Now using the expression of $D_{\alpha}$ in terms of the second moment 
$M_2$ (incorporating the unitary dimensional constant $\lambda$ in $w$) 
we define the reduced time variable
\beq
\widetilde{\tau}   \equiv {{(1+a)}\, \frac{M_2\,\tau}{w^{2\alpha}}} \,.
\eeq

\paragraph{The general spectrum.} With this definition, we rewrite \Ex{J_chi} as 
\beq
{\cal{J}}({\widetilde{\tau}}) = \, \frac{\alpha}{\alpha -1}\, 
e^{-\frac{\alpha}{\alpha-1}\frac{\left(2\pi\right)^{\alpha-1}}{\widetilde{\tau}}}\,%
 \int_{0}^{1} \chi ^{\frac{1}{\alpha - 1}}\, %
 e^{\frac{\alpha}{\alpha-1}\frac{\left(2\pi\right)^{\alpha-1}}{\widetilde{\tau}}\chi}\,d\chi \,. 
 \label{J_tau}
 \eeq
This is the general expression for the correlation spectrum obtained from the 
nonlinear theory of anomalous diffusion.  
 
\paragraph{The long time behavior.}   Observing that large $\tau$ in \Ex{defK} implies 
large $K$, the long time approximation of \Ex{J_chi}  gives
\beq
 {\cal{J}}(\tau) \approx %
 \frac{\alpha}{\alpha - 1}\, \frac{1}{K} \left(1 - e^{-K} \right) \,,
\eeq 
 or
\beq 
{\cal{J}}({\widetilde{\tau}})\approx %
   \left(\frac{\left(2\pi\right)^{\alpha-1}}{\widetilde{\tau}}\right)^{1/\alpha} %
 \left(1- e^{-\frac{\alpha}{\alpha-1}\frac{\left(2\pi\right)^{\alpha-1}}{\widetilde{\tau}}}\right) \,.
\label{LongCalJ}
\eeq

\paragraph{The classical spectrum.} For $a=1$ and in the limit $\alpha \longrightarrow 1$, 
we retrieve the result \Ex{GausSpec} for the Gaussian distribution 
\begin{equation}
\lim_{\alpha\rightarrow1}{\cal{J}}({\widetilde{\tau}}) = %
{\cal{J}}_{1}({\widetilde{\tau_1}})=\left(  1+\widetilde{\tau_1}\right)^{-1} %
\;\;\;\;\mbox{with}\;\;\;\;
\widetilde{\tau_1}   = { \frac{ 2 \,M_2\,\tau}{w^{2}}} \,.
\end{equation}

As an application of the theory we compare the theoretical  correlation spectrum with
fluorescence correlation experiments reported in \cite{Webb}. The results
obtained for the diffusion of fluorescently labeled lipid molecules in cell membranes
clearly show deviations from two-dimensional classical Brownian motion.  
In the absence of numerical data,  we processed the signal images in \cite{Webb} 
to obtain the data shown in fig.1 where they are
compared with our analytical results. While obviously the data are very poorly fit by the 
classical spectrum \Ex{GausSpec}, we find that the sub-diffusive nonlinear  correlation 
spectrum \Ex{J_tau} reproduces very well the experimental data indicating sub-diffusive 
molecular motion of lipid molecules in the cell membrane. 
\footnote{The data analysis in terms of sub-diffusion presented
in {\cite{Webb}} shows good agreement between the experimental results and the
theoretical correlation spectrum; however the analytical expression used to compute
the spectrum, Eq.(5) in {\cite{Webb}}, follows from the erroneous mere replacement 
of the Brownian mean squared displacement $4Dt$ by $\Gamma t^{\gamma}$ in the
expression of the classical spectrum (Eq.(3) in {\cite{Webb}} and \Ex{GausSpec} in the
present paper).}

 \begin{figure}
\resizebox{9cm}{8cm}{
\includegraphics{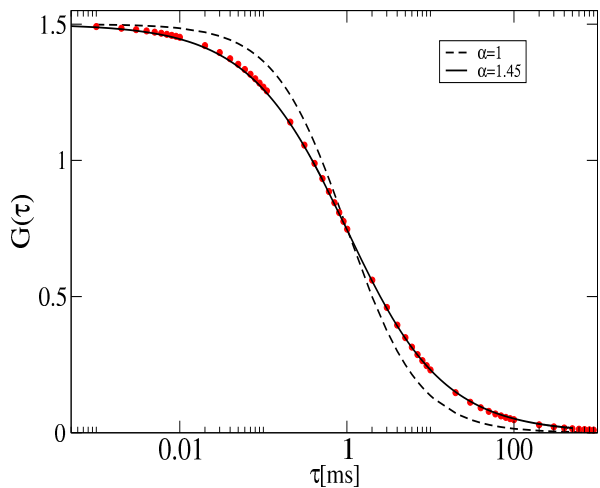}}
 \caption{Fluorescence correlation spectrum. 
 Experimental data (black dots) from Schwille {\it et al}, Fig.X  in \cite{Webb}, 
 showing the fluorescence correlation intensity of lipid molecules diffusing in
 the plasma membrane of rat cells (vertical axis: intensity normalized values; 
 horizontal axis: time in a.u.). The solid curve is the best-fit of the theoretical 
 spectrum \Ex{J_tau}. For comparison the dashed curve shows the best-fit 
 Gaussian profile \Ex{GausSpec}. %
}
 \label{fig_1}      
 \end{figure}

\begin{acknowledgements}
This work was supported in part by the European Space
Agency under contract number~ESA AO-2004-070.
\end{acknowledgements}



\end{document}